\documentclass[reprint,aps,prl,showpacs,twocolumn]{revtex4-1}%
\usepackage{amsmath}
\usepackage{amsfonts}
\usepackage{amssymb}
\usepackage{graphicx}
\usepackage[colorlinks,linkcolor=blue,anchorcolor=blue,citecolor=blue,urlcolor=black]%
{hyperref}
\usepackage{mathrsfs}
\usepackage{dcolumn}
\usepackage{bm}
\usepackage{epsfig}%
\def\be{\begin{equation}}
\def\ee{\end{equation}}
\begin{document}
\title{Coherent polariton dynamics in coupled highly-dissipative cavity quantum electrodynamics}
\author{Yong-Chun Liu$^{1,2}$}
\author{Xingsheng Luan$^{2}$}
\author{Hao-Kun Li$^{1}$}
\author{Qihuang Gong$^{1}$}
\author{Chee Wei Wong$^{2}$}
\email{cww2014@columbia.edu}
\author{Yun-Feng Xiao$^{1}$}
\email{Corresponding author: yfxiao@pku.edu.cn}
\altaffiliation{URL: www.phy.pku.edu.cn/$\sim$yfxiao/index.html}

\affiliation{$^{1}$State Key Laboratory for Mesoscopic Physics and School of Physics,
Peking University; Collaborative Innovation Center of Quantum Matter, Beijing
100871, P. R. China}
\affiliation{$^{2}$Optical Nanostructures Laboratory, Columbia University, New York, NY
10027, USA}
\date{\today}

\begin{abstract}
Coherent light-matter interaction at the single photon and electronic qubit
level promises the remarkable potential for nonclassical information
processing. Against the efforts of improving the figure of merit of the
cavities, here we demonstrate strong anharmonicity in the polariton dressed
states via dark state resonances in a highly dissipative cavity. It is shown
that vacuum Rabi oscillation occurs for a single quantum emitter inside a
cavity even with bosonic decay-to-interaction rate ratio exceeding $10^{2}$,
when the photon field is coupled to an auxiliary high-$Q$ cavity. Moreover,
photon blockade is observable in such a highly-dissipative cavity quantum
electrodynamics system. This study provides a promising platform for
overcoming decoherence and advancing the coherent manipulation of polariton qubits.

\end{abstract}

\pacs{42.50.Pq, 42.50.Ct}
\maketitle

Cavity quantum electrodynamics (QED) (for a review, see \cite{CQED06NatPhys})
provides a critical resource for quantum information processing
\cite{QE,sqc,sqc2,dqn,dqn2,haroche07,winelandPRL95,Waks12,WaksPRL06,Liu12,PRA12}
%xiaoAPL07
. For coherent manipulation, a key prerequisite is to reach the strong
coupling regime, where the emitter-field coupling strength exceeds the decay
rates of the emitter and the cavity field. In the past two decades great
efforts have been made to improve the quality ($Q$) factor and reduce the mode
volume ($V$) of the resonators for stronger interactions, using
Fabry-P\'{e}rot cavities \cite{FP92PRL,SCFP05NatBlock}, Bragg cavities
\cite{Pillar04Nat,Pillar07PRL,Pillar10NatMat}, whispering-gallery mode
cavities
\cite{SCWGM05PRL,SCWGM06Nat,SCWGM06NL,SCWGM06NL-2,SCWGM07Nat,SCWGM08NL} ,
photonic crystal cavities
\cite{PC04Nat,PC05SCI,PC07Nat,PC07Nat-2,PC08NatPhys,PC10NatPhys,PC12PRL},
hybrid plasmonic-photonic cavities \cite{Xiao12PRA} and transmission-line
microwave cavities \cite{circuit04Nat}, along with theoretical studies of
coupled-cavity QED through a waveguide
\cite{CoupledCQED07PRL,CoupledCQED07PRAzqyin,CoupledCQED07PRA,PRB12}. However,
it remains difficult to achieve high $Q$ and small $V$ simultaneously for the
same-type resonator. Fundamentally, this is related to the diffraction limit.
A smaller $V$ corresponds to a larger radiative decay rate and more
significant roughness scattering, leading to a lower $Q$. Different-type
resonators possess their own unique properties, but the trade-off between high
$Q$ and small $V$ still exists. For example, whispering-gallery mode cavities
possess ultrahigh $Q$ factors, while the mode volumes are relatively large;
for photonic crystal cavities, sub-wavelength light confinement can be
realized whereas the $Q$ factors are relatively low.

Unlike the efforts to improve the $Q/\sqrt{V}$ figure of merit of the
cavities, here we propose to reach the strong coupling regime via dark state
resonances, which removes the requirement for high $Q$ and small $V$ for the
same cavity. By coupling the originally weak-coupled cavity QED system with
high cavity dissipation to an auxiliary cavity mode with high-$Q$ but large
$V$, a strong dark state interaction takes place. We demonstrate that vacuum
Rabi oscillations and anharmonicity in the polariton dressed states occur even
when the cavity decay rate is two orders of magnitude larger than the
interaction rate.

\begin{figure}[tb]
%\begin{center}
\centerline{\includegraphics[width=7.5cm]{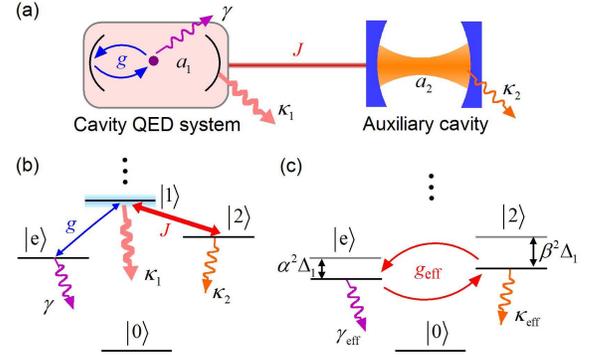}}
%7.5cm \columnwidth
%\end{center}
\caption{(color online) (a) Schematic of the cavity QED system coupled to an
auxiliary cavity. (b) Energy level diagram of the coupled system. The lowest
four energy levels are plotted, including the ground state $\left\vert
0\right\rangle $, the first excited state triplets $\left\vert \mathrm{e}%
\right\rangle $, $\left\vert 1\right\rangle $ and $\left\vert 2\right\rangle
$, which denote the states $|\mathrm{g}\rangle|0\rangle_{1}|0\rangle_{2}$,
$|\mathrm{e}\rangle|0\rangle_{1}|0\rangle_{2}$, $|\mathrm{g}\rangle
|1\rangle_{1}|0\rangle_{2}$ and $|\mathrm{g}\rangle|0\rangle_{1}|1\rangle_{2}%
$. (c) Sketch of the dark state interaction after eliminating state
$\left\vert 1\right\rangle $.}%
\label{Fig1}%
\end{figure}

As shown in Fig. \ref{Fig1}(a), a cavity QED system, consisting of a dipole
quantum emitter and a cavity, is coupled to an auxiliary cavity through a
short-length single-mode waveguide. Here we take Fabry-P\'{e}rot cavity QED
system as an example, while it allows generalization to other physical
implementations, including solid-state circuit QED systems. In the frame
rotating at the emitter's resonance frequency ${\omega_{\mathrm{e}}}$, the
system Hamiltonian reads $H={\Delta_{\mathrm{1}}a_{\mathrm{1}}^{\dag}%
a}_{\mathrm{1}}+{\Delta_{\mathrm{2}}a_{\mathrm{2}}^{\dag}a}_{\mathrm{2}}%
+{g{(}a_{\mathrm{1}}^{\dag}{\sigma}_{\mathrm{-}}+a}_{\mathrm{1}}{{\sigma
}_{\mathrm{+}}{)}}+{J{(a_{\mathrm{1}}^{\dag}a}_{\mathrm{2}}}+{{a_{\mathrm{2}%
}^{\dag}a}_{\mathrm{1}})}${, }where ${a}_{\mathrm{1}}$, ${a}_{\mathrm{2}}$ are
the annihilation operators of the two cavity modes; ${{\sigma}_{\mathrm{-}%
}\equiv}$ $|\mathrm{g}\rangle\langle\mathrm{e}|={{\sigma}_{\mathrm{+}}^{\dag}%
}$ stands for the descending operator of the emitter with $|\mathrm{g}\rangle$
($|\mathrm{e}\rangle$) being the ground (excited) state; ${\Delta_{\mathrm{1}%
}\equiv\omega_{\mathrm{1}}-\omega_{\mathrm{e}}}$ and ${\Delta_{\mathrm{2}%
}\equiv\omega_{\mathrm{2}}-\omega_{\mathrm{e}}}$ represent the detunings with
${\omega_{\mathrm{1}}}$ (${\omega_{\mathrm{2}}}$) being the resonance
frequency of mode ${a}_{\mathrm{1}}$ (${a}_{\mathrm{2}}$); $g$ denotes the
emitter-field coupling strength between the emitter and mode ${a}_{\mathrm{1}%
}$; $J$ describes the inter-cavity coupling strength between mode
${a}_{\mathrm{1}}$ and ${a}_{\mathrm{2}}$
\cite{CoupledNatPhoton12,CoupledPRA08,CoupledPRA10}. Without loss of
generality, we have assumed $g$ and $J$ to be real numbers. Taking the
dissipations into consideration, the system is described by the quantum master
equation $\dot{\rho}=i[\rho,H{]}+\kappa_{\mathrm{1}}\mathcal{D}[{a}%
_{\mathrm{1}}]\rho+\kappa_{\mathrm{2}}\mathcal{D}[{a}_{\mathrm{2}}]\rho
+\gamma\mathcal{D}[{{\sigma}_{\mathrm{-}}}]\rho$, where $\mathcal{D}[\hat
{o}]\rho=\hat{o}\rho\hat{o}^{\dag}{{-(\hat{o}^{\dag}\hat{o}\rho+\rho\hat
{o}^{\dag}\hat{o})/2}}$ is the standard dissipator in Lindblad form;
$\kappa_{\mathrm{1}}$, $\kappa_{\mathrm{2}}$ and $\gamma$ represent the decay
rates of modes ${a}_{\mathrm{1}}$, ${a}_{\mathrm{2}}$ and the emitter.

We show how highly dissipative cavity QED systems ($\kappa_{\mathrm{1}}\gg g$)
can be turned into the effective strong coupling regime via dark state
interaction. By eliminating mode ${a}_{\mathrm{1}}$, we obtain the effective
interaction between the emitter and the auxiliary cavity mode ${a}%
_{\mathrm{2}}$, with the effective Hamiltonian
\begin{equation}
H_{\mathrm{eff}}=\left(  {\Delta_{\mathrm{2}}-\beta^{2}\Delta_{\mathrm{1}}%
}\right)  {a_{\mathrm{2}}^{\dag}a}_{\mathrm{2}}-\frac{1}{2}\alpha^{2}%
\Delta_{\mathrm{1}}\sigma_{z}+{{{g}}}_{\mathrm{eff}}{{(}a_{\mathrm{2}}^{\dag
}{\sigma}_{\mathrm{-}}+a}_{\mathrm{2}}{{\sigma}_{\mathrm{+}}{)},} \label{Heff}%
\end{equation}
where $\sigma_{z}\equiv|\mathrm{e}\rangle\langle\mathrm{e}|-|\mathrm{g}%
\rangle\langle\mathrm{g}|$, $\alpha$ and $\beta$ represent the scaled
dimensionless interaction parameters given by $\alpha=g/(\Delta_{\mathrm{1}%
}^{2}+\kappa_{\mathrm{1}}^{2}/4)^{1/2}\ $and $\beta=J/(\Delta_{\mathrm{1}}%
^{2}+\kappa_{\mathrm{1}}^{2}/4)^{1/2}$, respectively. The effective coupling
strength, detuning, decay rates of the cavity field and the emitter are
described by
\begin{align}
{g}_{\mathrm{eff}}  &  =\beta g,\text{ \ }\Delta_{\mathrm{eff}}=\Delta
_{\mathrm{2}}+\left(  \alpha^{2}-\beta^{2}\right)  \Delta_{\mathrm{1}%
},\nonumber\\
\kappa_{\mathrm{eff}}  &  =\kappa_{\mathrm{2}}+\beta^{2}\kappa_{\mathrm{1}%
},\text{ \ }\gamma_{\mathrm{eff}}=\gamma+\alpha^{2}\kappa_{\mathrm{1}}.
\label{eff}%
\end{align}

In Fig. \ref{Fig1}(b) we plot the energy level diagram, which displays the
lowest four energy levels of the system. It reveals that the emitter-field
interaction between state $|\mathrm{e}\rangle$ (short for $|\mathrm{e}%
\rangle|0\rangle_{1}|0\rangle_{2}$) and state $|1\rangle$ (short for
$|\mathrm{g}\rangle|1\rangle_{1}|0\rangle_{2}$), together with the
inter-cavity interaction between state $|2\rangle$ (short for $|\mathrm{g}%
\rangle|0\rangle_{1}|1\rangle_{2}$) and state $|1\rangle$, yields the
effective dark state interaction between state $|\mathrm{e}\rangle$ and state
$|2\rangle$. As shown in Eqs. (\ref{Heff})-(\ref{eff}) and illustrated in Fig.
\ref{Fig1}(c), after the the elimination of state states $|1\rangle$, the
states $|\mathrm{e}\rangle$ and $|2\rangle$ acquire energy shifts of
$-\alpha^{2}\Delta_{\mathrm{1}}$ and $-\beta^{2}\Delta_{\mathrm{1}}$, together
with broadenings of $\alpha^{2}\kappa_{\mathrm{1}}$ and $\beta^{2}%
\kappa_{\mathrm{1}}$.

\begin{figure}[tb]
%\begin{center}
\centerline{\includegraphics[width=\columnwidth]{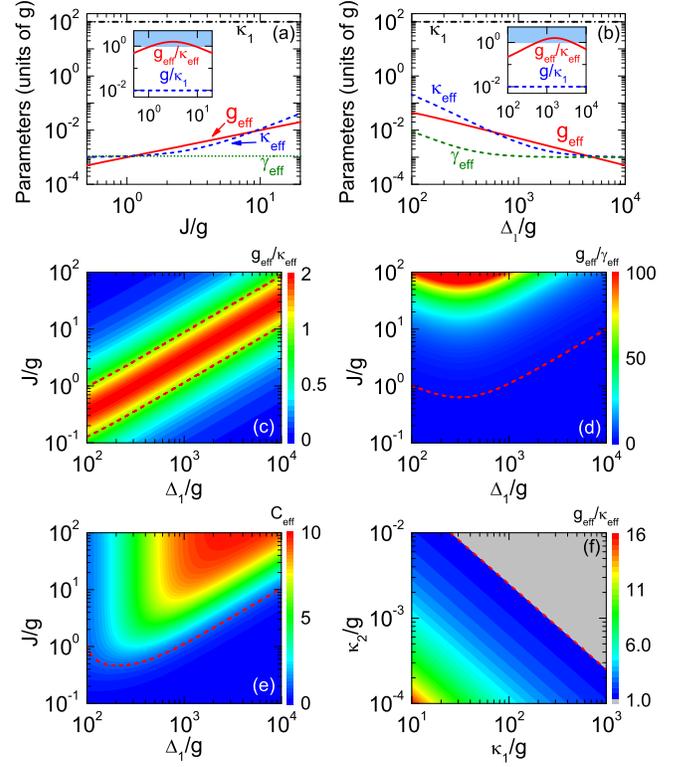}}
%7.5cm \columnwidth
%\end{center}
\caption{(color online) Parameters $\kappa_{\mathrm{1}}$ (black dashed-dotted
curves), $g_{\mathrm{eff}}$ (red solid curves), $\kappa_{\mathrm{eff}}$ (blue
dashed curves) and $\gamma_{\mathrm{eff}}$ (green dotted curves) as functions
of inter-cavity interaction strength (a) and the first cavity mode's detuning
(b). Both the horizontal and vertical axes are in the units of the
emitter-field coupling strength $g$. The insets in (a) and (b) show
$g_{\mathrm{eff}}/\kappa_{\mathrm{eff}}$ (red solid curves) and $g/\kappa
_{\mathrm{1}}$ (blue dashed curves); the blue shaded regions indicate
$g_{\mathrm{eff}}/\kappa_{\mathrm{eff}}>1$. (c)-(e): Contour plots of
$g_{\mathrm{eff}}/\kappa_{\mathrm{eff}}$, $g_{\mathrm{eff}}/\gamma
_{\mathrm{eff}}$ and the cooperativity $C_{\mathrm{eff}}$ as functions of
$\Delta_{\mathrm{1}}/g$ and $J/g$; the red dashed curves denote the contour
value of $1$. In (a), $\Delta_{\mathrm{1}}/\kappa_{\mathrm{1}}=10$; in (b),
$J/g=5$; in (a)-(e), $\kappa_{\mathrm{1}}/g=100$, $\kappa_{\mathrm{2}%
}/g=10^{-3}$,$\ \gamma/g=10^{-3}$ and $\Delta_{\mathrm{2}}=(\beta^{2}%
-\alpha^{2})\Delta_{\mathrm{1}}$. (f) Contour plot of $g_{\mathrm{eff}}%
/\kappa_{\mathrm{eff}}$ as functions of $\kappa_{\mathrm{1}}/g$ and
$\kappa_{\mathrm{2}}/g$ for $\beta=\sqrt{\kappa_{\mathrm{2}}/\kappa
_{\mathrm{1}}}$; the red dashed curve denotes $\kappa_{\mathrm{2}}%
=g^{2}/(4\kappa_{\mathrm{1}})$.}%
\label{Fig2}%
\end{figure}

Equations (\ref{eff}) show that the effective coupling strength ${g}%
_{\mathrm{eff}}$ depends linearly on $\beta$ while the effective decay rates
$\kappa_{\mathrm{eff}}$ and $\gamma_{\mathrm{eff}}$ are quadratic functions of
$\beta$ and $\alpha$, respectively. As a result, for $\left(  \alpha
,\beta\right)  \ll1$, the effective coupling strength will be larger than the
decay rates, driving the effective interaction into the strong coupling
regime. In Figs. \ref{Fig2}(a) and (b) the parameters given by Eq. (\ref{eff})
as functions of inter-cavity interaction strength $J$ and the first cavity
mode's detuning $\Delta_{\mathrm{1}}$ are plotted, respectively. It reveals
that with a suitable $J$ and $\Delta_{\mathrm{1}}$, the effective coupling
strength ${{{g}}}_{\mathrm{eff}}$ exceeds both decay rates $\kappa
_{\mathrm{eff}}$ and $\gamma_{\mathrm{eff}}$, even for large cavity decay rate
$\kappa_{\mathrm{1}}/g=100$. As shown in the insets of Figs. \ref{Fig2}(a) and
(b), the ranges of $J$ and $\Delta_{\mathrm{1}}$ for effective strong coupling
have both lower and upper bounds. To gain more insights on the parameter ranges, in Figs.
\ref{Fig2}(c)-(e) we plot ${g}_{\mathrm{eff}}/\kappa_{\mathrm{eff}}$,
${g}_{\mathrm{eff}}/\gamma_{\mathrm{eff}}$ and the cooperativity parameter
$C_{\mathrm{eff}}\equiv{g}_{\mathrm{eff}}^{2}/(\kappa_{\mathrm{eff}}%
\gamma_{\mathrm{eff}})$ as functions of $\Delta_{\mathrm{1}}$ and $J$. It
reveals that a large $\Delta_{\mathrm{1}}$ and a corresponding large $J$ lead
the system deeply into the effective strong coupling regime. Examining Eq.
(\ref{eff}), for $J>g>\kappa_{\mathrm{2}}\sim\gamma$, it gives $\kappa
_{\mathrm{eff}}>\gamma_{\mathrm{eff}}$ with negligible $\gamma_{\mathrm{eff}}%
$. In this case, the maximum effective coupling-to-decay rate ratio reads
${g}_{\mathrm{eff}}/\kappa_{\mathrm{eff}}=g/(2\sqrt{\kappa_{\mathrm{1}}%
\kappa_{\mathrm{2}}})$, obtained when $\beta=\sqrt{\kappa_{\mathrm{2}}%
/\kappa_{\mathrm{1}}}$. Thus the strong coupling condition ${g}_{\mathrm{eff}%
}>\kappa_{\mathrm{eff}}$ can be fulfilled when $\kappa_{\mathrm{2}}%
<g^{2}/(4\kappa_{\mathrm{1}})$.\ This is verified by the contour plot in Fig.
\ref{Fig2}(f), which displays ${g}_{\mathrm{eff}}/\kappa_{\mathrm{eff}}$ as a
function of $\kappa_{\mathrm{1}}$ and $\kappa_{\mathrm{2}}$. The bottom left
region indicates the strong effective coupling parameter regime, with
${g}_{\mathrm{eff}}$ in excess of $\kappa_{\mathrm{eff}}$ by more than one
order of magnitude.

\begin{figure}[tb]
%\begin{center}
\centerline{\includegraphics[width=7.5cm]{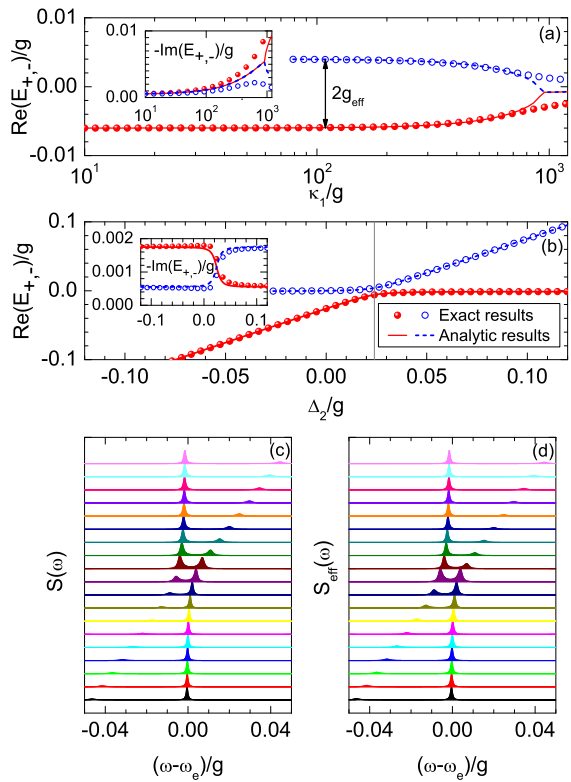}}
%7.5cm \columnwidth
%\end{center}
\caption{(color online) (a) and (b): Eigenvalues $E_{\pm}$ for states
$|0\rangle_{1}|1,\pm\rangle_{\mathrm{e}2}$ as functions of $\kappa
_{\mathrm{1}}/g$ and $\Delta_{\mathrm{2}}/g$. The main figures and the insets
show the real and imaginary parts of the eigenvalues, respectively. The
circles correspond to the exact results and the curves denote the results
obtained from the effective Hamiltonian and effective parameters [Eqs.
(\ref{Heff})-(\ref{eff})]. The gray vertical line in (b) denotes
${\Delta_{\mathrm{2}}}=(\beta^{2}-\alpha^{2})\Delta_{\mathrm{1}}$. (c) and
(d): Normalized spectra $S(\omega)$ and effective spectra $S_{\mathrm{eff}%
}(\omega)$ of the emitter for various ${\Delta_{\mathrm{2}}}$. From top to
bottom, ${\Delta_{\mathrm{2}}}$ decrease from $(\beta^{2}-\alpha^{2}%
)\Delta_{\mathrm{1}}-9g_{\mathrm{eff}}$ to $(\beta^{2}-\alpha^{2}%
)\Delta_{\mathrm{1}}+9g_{\mathrm{eff}}$ with step $g_{\mathrm{eff}}$. The
common parameters are the same as Fig. \ref{Fig2}(a)-(e). }%
\label{Fig3}%
\end{figure}

To demonstrate that the effective parameters in Eq. (\ref{eff}) exactly
describe the physical interaction, we diagonalize the system Hamiltonian in
the subspace of the first excited states. Using the non-Hermitian Hamiltonian where the decays
are taken into account, the eigenenergies and the broadenings of each states
are obtained as the real and imaginary parts of the eigenvalues, respectively.
For the first excited states, after the diagonalization under large detuning
$\Delta_{\mathrm{1}}$, the eigenstates read $|1\rangle_{1}|0\rangle
_{\mathrm{e}2}\simeq|\mathrm{g}\rangle|1\rangle_{1}|0\rangle_{2}$,
$|0\rangle_{1}|1,\pm\rangle_{\mathrm{e}2}\simeq(|\mathrm{e}\rangle
|0\rangle_{1}|0\rangle_{2}\pm|\mathrm{g}\rangle|0\rangle_{1}|1\rangle
_{2})/\sqrt{2}$. It reveals that the states $|0\rangle_{1}|1,\pm
\rangle_{\mathrm{e}2}$ are \textit{dark state doublets} with respect to the
decay of mode ${a}_{\mathrm{1}}$. In Figs. \ref{Fig3}(a) we plot the real and
imaginary parts of the eigenvalues $E_{\pm}$ for the dark state doublets
$|0\rangle_{1}|1,\pm\rangle_{\mathrm{e}2}$ as functions of $\kappa
_{\mathrm{1}}/g$, where the real (imaginary) parts represent the eigenenergies
(linewidths) of the states. It shows that the eigenenergies of the two states
are split by $2{{{g}}}_{\mathrm{eff}}=0.01g$, and the linewidths are much
smaller than the energy splitting (inset), even for $\kappa_{\mathrm{1}}/g$
exceeding $100$. Note that the global energy shift of $0.001g$ ($=-\alpha
^{2}\Delta_{\mathrm{1}}$) can be eliminated by applying a unitary
transformation to the effective Hamiltonian [Eq. (\ref{Heff})]. The results
obtained from the effective Hamiltonian and effective parameters [Eqs.
(\ref{Heff}) and (\ref{eff})] are in good accordance with the exact results
for both the real and imaginary parts of the eigenvalues. For $\kappa
_{\mathrm{1}}/g\gtrsim800$, discrepancy occurs because $\Delta_{\mathrm{1}}%
\gg\kappa_{\mathrm{1}}$ is not satisfied.

In Fig. \ref{Fig3}(b) we plot the eigenenergies and linewidths for states
$|0\rangle_{1}|1,\pm\rangle_{\mathrm{e}2}$ as functions of the detuning
between mode ${a}_{\mathrm{2}}$ and the emitter ($\Delta_{\mathrm{2}}/g$).
Prominent avoided crossing phenomenon occurs for the eigenenergies, which
occurs for the effective resonant case $\Delta_{\mathrm{eff}}=0$ (gray
vertical line). Near the avoided crossing point the linewidths of the two
polariton states are averaged compared with the large $\Delta_{\mathrm{eff}}$
case (inset), and are swapped for increasing detuning as indication of the
quantum strong coupling. The avoided crossing is further examined in Fig.
\ref{Fig3}(c), which shows the emitter's spectra $S(\omega)$ for various detunings
$\Delta_{\mathrm{2}}$ through the weak excitation of mode ${a}_{\mathrm{2}}$.
It shows close agreement with the effective spectra $S_{\mathrm{eff}}(\omega)$
obtained from the effective interaction [Fig. \ref{Fig3}(d)].

\begin{figure}[tb]
%\begin{center}
\centerline{\includegraphics[width=\columnwidth]{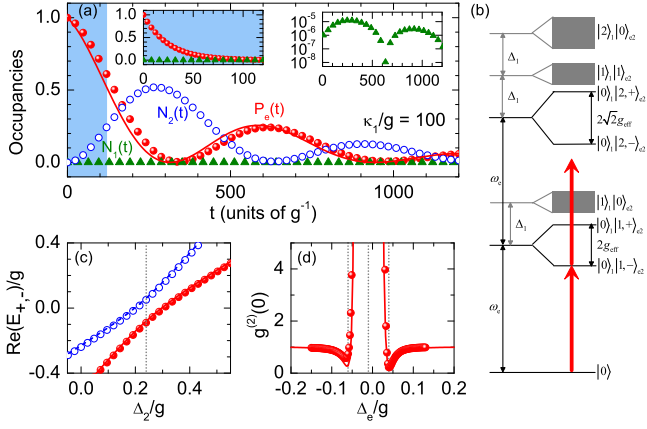}}
%7.5cm \columnwidth
%\end{center}
\caption{ (color online) (a) Time evolution of the mean photon numbers
$N_{1}(t)$ (green triangles), $N_{2}(t)$ (blue open circles) and the
probability for the emitter being in the excited state $P_{\mathrm{e}}(t)$
(red closed circle) for $\kappa_{\mathrm{1}}/g=100$. The red solid curves
correspond to the analytical results of $P_{\mathrm{e}}^{\mathrm{eff}}(t)$
[Eq. (\ref{Peff})]. Left inset: Comparative $N_{1}(t)$ (green triangles) and
$P_{\mathrm{e}}(t)$ (red closed circles) without the auxiliary cavity ($J=0$)
and for $\Delta_{\mathrm{1}}=0$; the range of the horizontal axis is the same
as the shaded region of the main figure. Right inset: Log scale plot of
$N_{1}(t)$. (b) Energy level diagram of the ground state, the first and second
excited states for interpretation of the photon blockade effect. (c)
Eigenenergies of the second excited states $|0\rangle_{1}|2,\pm\rangle
_{\mathrm{e}2}$ as functions of ${\Delta_{\mathrm{2}}}/g$ for $\kappa
_{\mathrm{1}}/g=10$. The gray vertical line denotes ${\Delta_{\mathrm{2}}%
}=(\beta^{2}-\alpha^{2})\Delta_{\mathrm{1}}$. (d) Second-order correlation
function $g^{(2)}(0)$ as a function of the probe-emitter detuning
$\Delta_{\mathrm{e}}$ for $\kappa_{\mathrm{1}}/g=10$. The gray vertical line
denotes $\Delta_{\mathrm{e}}=-\alpha^{2}\Delta_{\mathrm{1}}-g_{\mathrm{eff}}$,
$-\alpha^{2}\Delta_{\mathrm{1}}$ and $-\alpha^{2}\Delta_{\mathrm{1}%
}+g_{\mathrm{eff}}$ (from left to right). The circles correspond to the exact
results and the curves indicate the results obtained from the effective
Hamiltonian and effective parameters [Eqs. (\ref{Heff}) and (\ref{eff})]. The
common parameters are the same as Fig. \ref{Fig2}(a)-(e). }%
\label{Fig4}%
\end{figure}

In the time domain, vacuum Rabi oscillation is a direct evidence of the
coherent energy exchange between the emitter and the cavity photon field. Here
we numerically solve the quantum master equation to obtain the exact results.
We assume initially the emitter is in the excited state and the two cavity
modes are in their vacuum states, then we obtain the exact numerical results
for the time evolution of the mean photon numbers $N_{1}(t)=\langle
{a_{\mathrm{1}}^{\dag}a}_{\mathrm{1}}\rangle$, $N_{2}(t)=\langle
{a_{\mathrm{2}}^{\dag}a}_{\mathrm{2}}\rangle$ and the probability for the
emitter being in the excited state $P_{\mathrm{e}}(t)=(\langle\sigma
_{z}\rangle+1)/2$. As shown in Fig. \ref{Fig4}(a), even for $\kappa
_{\mathrm{1}}/g=100$, vacuum Rabi oscillation phenomenon occurs for several
periods, revealing that the decoherence time is much longer than the energy
exchange period. This is in contrary to the case without the auxiliary cavity
as shown in the left inset of Fig. \ref{Fig4}(a), where the emitter
exponentially decays from the excited state. Note that the occupancy in mode
${a}_{\mathrm{1}}$ oscillates with the maximum photon number below $10^{-5}$
as shown in the right inset of Fig. \ref{Fig4}(a)], while the occupancy in
mode ${a}_{\mathrm{2}}$ oscillates with the maximum photon number exceeding
0.5. This reveals that the interaction is mainly between the emitter and mode
${a}_{\mathrm{2}}$, while mode ${a}_{\mathrm{1}}$ is only virtually excited.
The analytical results for the emitter's occupancy in the excited state,
obtained from the effective parameters [Eq. (\ref{eff})], is described by
\begin{equation}
P_{\mathrm{e}}^{\mathrm{eff}}(t)=\exp(-\frac{\kappa_{\mathrm{eff}}%
+\gamma_{\mathrm{eff}}}{2}t)\cos^{2}(g_{\mathrm{eff}}t). \label{Peff}%
\end{equation}
With vacuum Rabi frequencies $\Omega_{\mathrm{R}}=2g_{\mathrm{eff}}$ and the
decay rates $(\kappa_{\mathrm{eff}}+\gamma_{\mathrm{eff}})/2$, the results in
the effective dark state picture (red solid curve) are in good accordance with
the exact numerical results (red closed circles).

The effective strong coupling offers great potential for single-photon
manipulation and quantum logic gate operation. For example, photon blockade
phenomenon \cite{Blockade05Nat,Blockade11PRL} occurs in this coupled system,
as illustrated in Fig. \ref{Fig4}(b), where the energy spectrum for the ground
state, the first and second excited states are plotted. The first excited
state has triplet sub-levels, and the second excited state has quintet
sub-levels including $|0\rangle_{1}|2,\pm\rangle_{\mathrm{e}2}$,
$|1\rangle_{1}|1,\pm\rangle_{\mathrm{e}2}$ and $|2\rangle_{1}|0\rangle
_{\mathrm{e}2}$. The computed energy levels $|0\rangle_{1}|2,\pm
\rangle_{\mathrm{e}2}$ are shown in Fig. \ref{Fig4}(c), which are dark state
doublets with energy splitting of $2\sqrt{2}{g}_{\mathrm{eff}}$ at the minimal
avoided crossing point.
Due to the strong anharmonicity of the level spacing between the polariton
dressed states, photon blockade of the second photon by the first photon can
occur. This is quantitatively characterized by the zero-delay second-order
correlation function $g^{(2)}(0)\equiv\lim_{t\rightarrow\infty}\langle
{a_{\mathrm{2}}^{\dag}a_{\mathrm{2}}^{\dag}a}_{\mathrm{2}}{a}_{\mathrm{2}%
}\rangle(t)/\langle{a_{\mathrm{2}}^{\dag}a}_{\mathrm{2}}\rangle^{2}(t)$. We
use a weak probe laser input with frequency $\omega$ to obtain the exact
results of $g^{(2)}(0)$ numerically. In Fig. \ref{Fig4}(d) $g^{(2)}(0)$ as a
function of the probe-emitter detuning $\Delta_{\mathrm{e}}\equiv\omega
-\omega_{\mathrm{e}}$ is plotted. It reveals that $g^{(2)}(0)\ $approaches $0$
for $\Delta_{\mathrm{e}}=-\alpha^{2}\Delta_{\mathrm{1}}\pm g_{\mathrm{eff}}$,
indicating strong antibunching effect and sub-Poissonian photon statistics.
Under such strong coupling regime, with an external field pumping the system,
it is also promising for the generation of one-atom lasing
\cite{Lasing92PRA,Lasing11PRA,Lasing03Nat}.

It should be noted that, although the auxiliary cavity is required to be
high-$Q$ ($\kappa_{\mathrm{2}}<g$), it does not need to interact directly with
the emitter, and its mode volume is not necessary to be small. Therefore, the
scheme does not require a high figure of merit $Q/\sqrt{V}$ for the auxiliary
cavity. Together with the allowed low $Q$ factor for the primary cavity, both
the two cavities can be low in figure of merit $Q/\sqrt{V}$. This approach is
also generic and can be applied to any cavity QED systems with different
physical implementations, including solid-state circuit QED systems. In the
viewpoint of mode density shaping \cite{Purcell10APL}, at the second cavity
mode's resonance frequency the system's mode density is enhanced, and it leads
to the effective interaction between the second cavity mode and the emitter.

In summary, we have presented a protocol for realizing effective strong
coupling in a highly-dissipative cavity QED system. By employing the coupled
cavity configuration, we show that a highly dissipative cavity interacting
simultaneously with a single emitter and an auxiliary cavity leads to the dark
state resonance between the emitter and the auxiliary cavity. It is
demonstrated that effective strong coupling can be achieved even with low
$Q/\sqrt{V}$ cavities, with prominent vacuum Rabi oscillation and ladder
anharmonicity phenomena for photon blockade. The cavity coupled to the emitter
can be highly dissipative even with the decay rate in excess of the
interaction strength by two orders of magnitude. The system enables single
photon manipulation like photon blockade and quantum logic gate operations.
This approach offers opportunities to exploit both theoretical and
experimental physics in the strong light-matter interaction regime without
stringent cavity requirements.

\begin{acknowledgments}
This work was supported by the 973 program (No. 2013CB921904, No.
2013CB328704), NSFC (Nos. 11004003, 11222440, and 11121091), RFDPH (No.
20120001110068), NSF MWN and IGERT awards (DMR-1108176 and DGE-1069240).
H.K.L. was supported by the National Fund for Fostering Talents of Basic
Science (Grants No. J1030310 and No. J1103205)
\end{acknowledgments}

\end{document}